\documentclass[12pt,epsf]{iopart}
\newcommand{\beq}{\begin{equation}}
\newcommand{\eeq}{\end{equation}}
\newcommand{\beqa}{\begin{eqnarray}}
\newcommand{\eeqa}{\end{eqnarray}}
\newcommand{\simg}
   {\mathrel{\raise.3ex\hbox{$>$\kern-.75em\lower1ex\hbox{$\sim$}}}}
\newcommand{\siml}
   {\mathrel{\raise.3ex\hbox{$<$\kern-.75em\lower1ex\hbox{$\sim$}}}}
\newcommand{\epm}{e^{\pm}}
\newcommand{\ps}{\rm Ps}
\newcommand{\ra}{\rightarrow}
\usepackage{graphicx}
\usepackage{amssymb,latexsym}

\begin{document}


\title[Ps and Dark Energy]{
Does Positronium Form in the Universe ?
}

\author{Takeshi Chiba\dag, Naoshi Sugiyama\ddag}
\address{\dag Department of Physics, 
College of Humanities and Sciences, Nihon University, Setagaya, Tokyo 156-8550, Japan}
\address{\ddag Division of Theoretical Astrophysics, National Astronomical Observatory,
Japan, Mitaka, Tokyo 181-8588, Japan
}

\begin{abstract}
Positronium (the bound state of electron and positron) has been thought to 
be formed after proton decay ($>10^{34}$yr) through collisional 
recombination and then decays by pair annihilation, thereby changing 
the matter content of the universe. 
We revisit the issue of the formation of positronium in the long-term
future of the universe in light of recent indication that the universe
is dominated by dark energy and dark matter. 
We find that if the equation of state of dark energy $w$ is less than $-1/3$  
(including the cosmological constant $w=-1$), then the
formation of positronium would not be possible, while it is possible 
through bound-bound transitions 
 for $-1/3\siml w\siml-0.2$, 
or through collisional recombination for $w\simg-0.2$. 
The radiation from $\epm$ pair annihilation cannot dominate over $\epm$, 
while that from proton decay will dominate over baryon and $\epm$ for a while 
but not over dark matter. 
\end{abstract}

\maketitle

\section{Introduction}

Since the pioneering work by Dyson \cite{dyson} (see also \cite{davies}), 
the physical processes that would occur in the far future of the universe
have been studied by several authors \cite{bt,pm,dicus,bt2,al}. 
One of interest 
then was the effect of proton decay on the matter content of the universe. 
It was found for an open universe that the decay of protons 
(with lifetime $t_d$) will produce nonthermal radiation which will 
dominate the universe from $t_d$ to $10t_d$ and then matter domination 
follows \cite{dicus}. Moreover, electrons and positrons from proton decay 
will bound to become positroniums and eventually annihilate \cite{bt}. 
However, the annihilation of positron-electron pair occurs so slowly that 
the radiation from pair annihilation cannot dominate the universe \cite{pm}.

Today the situation in cosmology becomes much changed. 
The new ingredients after Page and McKee's paper are: (1) Cold dark matter; baryons are not
 dominant component of the universe,  (2) Dark energy; the universe is presently 
dominated by energy density with negative pressure. 
As an additional minor point, we note that the lower-bound of 
the proton decay, $t_d> 10^{33}$yr \cite{pdg}, is much larger than the 
previous bound ($t_d> 10^{30}$yr). 
The purpose of this paper is to revisit the nature of the long-term future 
of the universe taking into account these new ingredients after 
Page and McKee's study \cite{pm}. Throughout this paper, we assume that the laws 
of physics continue to hold unchanged. In particular, we assume that the charge 
of electron does not vary with time.


\section{Positronium Formation in an Expanding Universe}

According to the grand unified theory (GUT) of elementary forces, 
all baryons are unstable. The life time of a baryon $t_d$ is predicted to be 
typically some $10^{34}$ years for (supersymmetric) GUT models. 
The positrons and electrons, when first produced from nucleon decay, 
will have energies of several hundred MeV so that they will appear 
as relativistic particles until the universe has expanded by a factor of 
the decay energy divided by the 
electron mass. Let $q_-$ be the electron mass divided by the average decay 
energy per the electron mass and $q_+$ be the same ratio for positrons. 
Let $f_-$ be the average number of electrons produced per nucleon decay and 
$f_+$ be the same quantity for positrons. For SU(5) GUT model,\footnote{We adopt  
non-supersymmetric SU(5) GUT as an example only for simplicity. Our qualitative results will not 
be sensitive to the detail of the model.}
 assuming that $16\%$ 
of the nucleons in the universe are neutrons, they are 
given by \cite{dicus,paul}
\beqa
&&q_-=5.9\times 10^{-3},~~~f_-=0.54\\ 
&&q_+=2.1\times 10^{-3},~~~f_+=1.38. 
\eeqa
Thus, baryon density $\rho_B$, $\epm$'s density $\rho_e$ and 
radiation density $\rho_r$ evolve according to \footnote{
Here we have assumed that all of the baryons in the universe are free at 
the epoch of proton decay. In reality, however, a large fraction of the 
baryons would be locked up in degenerate remnants (mostly in white dwarfs) at this future time. 
When protons decay within dwarfs, the positrons annihilate with electrons, and 
the resulting high energy photons are reprocessed into low energy photons. 
As a result, protons that are locked up in degenerate remnants are essentially lost 
to the processes discussed here. The subsequent analyses would apply to the baryons that 
will not be processed into stars and stellar remnants.\cite{adams,al}}\cite{dicus}
\beqa
\dot\rho_B&=&-3H\rho_B-t_d^{-1}\rho_B\\
\dot\rho_e&=&-3H\rho_e
+{m\rho_B\over m_pt_d}\left(q_-f_-\theta(t_- -t_0)+q_+f_+\theta(t_+ -t_0)
\right)\label{rhoedot}\\
\dot\rho_r&=&-4H\rho_r 
+t_d^{-1}\rho_B-{m\rho_B\over m_pt_d}\left(q_-f_-\theta(t_- -t_0)+
q_+f_+\theta(t_+ -t_0)
\right),\\
H^2&=&{8\pi G\over 3}(\rho_{DE}+\rho_{DM}+\rho_B+\rho_e;\rho_r)
\eeqa
where $m$ is the mass of electron(positron), $m_p$ is the proton mass.
$t_-(t_+)$ is the time when the universe expand $1/q_-(1/q_+)$ times from now 
($t_0$), and $\rho_{DE}$ ($\rho_{DM}$) is the energy density of dark energy (matter). 
The second term in Eq.(\ref{rhoedot}) includes the effect that 
electrons (and positrons) from proton decay are initially considered radiation 
and will become matter when the universe is $1/q_-$ ($1/q_+$) times expanded.
Assuming that the universe is dominated by dark energy with 
the equation-of-state $w$ and dark matter with $\rho_{DM}(t_0)=
6\times \rho_B(t_0)$ which is stable\footnote{We assume that dark matter is 
stable. The subsequent discussion will be greatly changed for unstable dark 
matter, which will be strongly model-dependent however.} 
and taking $\rho_r=0$ and 
$\rho_e=(m/m_p)\rho_B(t_0)(f_+-f_-)$ (from charge neutrality) 
at the present time $t=t_0$, we solve the above equations. 
The results are shown in Fig. 1 for $w=-0.3$ and $w=-1$. We find that radiation 
can dominate over baryon, electron and positron for only a short period 
after proton decay for $w=-0.3$. 

\begin{figure}
\includegraphics[width=\hsize]{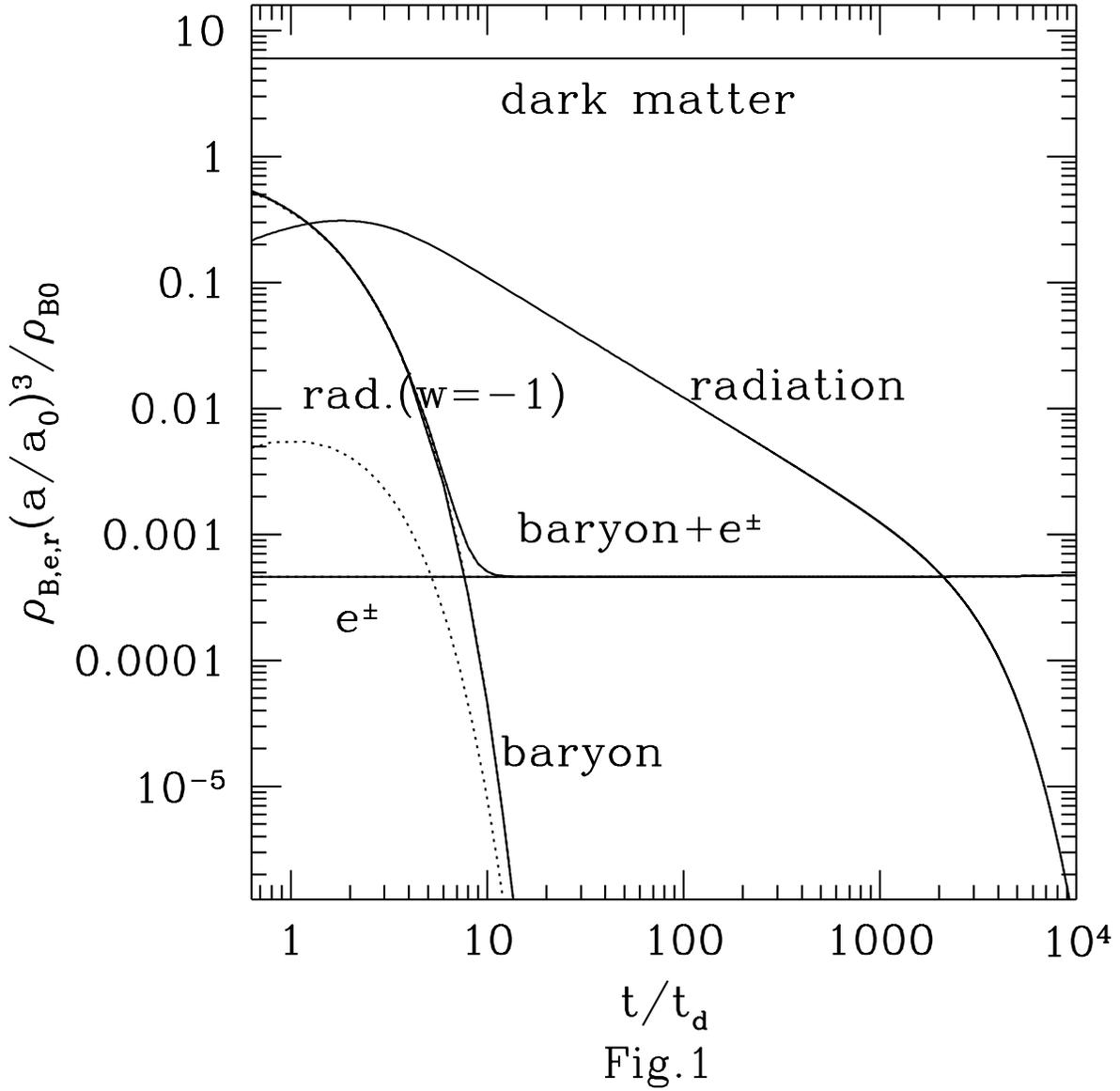}
\caption{
Evolution of $\rho_B$, $\rho_e$ and 
$\rho_r$ for $w=-0.3$ (solid lines) and for $w=-1$(dashed lines). $a_0$ corresponds to 
the scale factor at the present time $t_0$. $\rho_e$ does slightly increase after 
baryon decay, although it is barely seen in the figure.}
\end{figure}

\subsection{Necessary Condition}

When a baryon with mass $m_p$ decays, a significant fraction of its 
energy will go into $e^{\pm}$s (their mass $m$) with average rms momentum 
and energy $p=\gamma m$ at that time $t_d$, where $\gamma m$ is roughly 
the fraction of energy $m_p$ of each decay that goes into each $e^{\pm}$ 
produced, say $\gamma\sim m_p/2m\sim 10^3$ in accord with the previous 
example ($1/q_-,1/q_+$). As the universe expands, in the absence of 
non-gravitational interactions 
the rms momentum will redshift to $\gamma m(a_d/a)$. Once this becomes less than
$m$ (when $a>\gamma a_d$), the average kinetic energy per $\epm$ will be 
$p^2/2m\sim \gamma^2 m(a_d/a)^2$.  
In the following, we shall consider a cosmological model with dark energy with 
equation-of-state $w$. Hence the expansion law is $a\propto t^{2/3(1+w)}$.

As Barrow and Tipler argued, a positronium will form if the average 
energy of an electron $E$ at average distance $r$ from a positron 
\beq
E\sim \gamma^2 m(a_d/a)^2 -e^2/r
\label{energy}
\eeq
is negative \cite{bt}.\footnote{This is the case for bound-bound transition. Positronium can 
be formed much earlier through collisional recombination \cite{pm}. The detailed anaylsis 
is given in Appendix.}
 Note that the Coulomb energy decays slower than 
the kinetic energy and hence $E$ will be eventually negative. 
Here the mean separation $r$ is estimated from 
the number density of $\epm$, $N$
\beq
r\sim N^{-1/3}.
\eeq
$N$ is related to the number density of baryon at the decay time $N_d$ as
\beq
N\sim N_d(a_d/a)^3.
\eeq
$N_d$ is written in terms of the ratio of baryon number density to entropy 
density 
$n_B/s\sim 10^{-10}(\Omega_Bh^2/10^{-2})$ as
\beq
N_d\sim n_{B0}(a_0/a_d)^3\sim 10^{-7} \left(n_Bs^{-1}/10^{-10}\right)
(a_0/a_d)^3{\rm cm^{-3}},
\eeq
where $a_0$ is the scale factor at present. 
An electron-positron pair will bound if $E<0$ when $a>a_*$, where from Eq.
(\ref{energy})
\beq
a_*/a_d\sim {\gamma^2me^{-2}N_d^{-1/3}}\sim 10^{21}10^{16/(1+w)}
(t_d/10^{34}{\rm yr})^{2/3(1+w)}.
\label{cond1}
\eeq

In an accerelating universe ($w<-1/3$), 
another condition for the formation of positronium arises: repulsive 
gravitational force sourced by dark energy should not 
overcome the attractive Coulomb force: 
\beq
e^2/r^2 > -mr\ddot a/a=-mr(1+3w)H^2/6. 
\label{bound}
\eeq
The question is whether Eq.(\ref{bound}) is satisfied at that time. 
{}From
\beq
r\sim {e^2\gamma^{-2} m^{-1}}(a_*/a_d)^2
\eeq
and
\beq
H^{-1}\sim H_0^{-1}(a/a_0)^{3(1+w)/2}\sim 10^{34}
(t_d/10^{34}{\rm yr})(a/a_d)^{3(1+w)/2}{\rm yr},
\eeq
it is required that (for $w<-1/3$)
\beq
a_*/a_d<10^{49/(1-w)}(t_d/10^{34}{\rm yr})^{2/3(1-w)},
\label{cond2}
\eeq
where we have neglected an unimportant numerical coefficient coming 
from $|1+3w|$ in Eq.(\ref{bound}). 
Two conditions Eq.(\ref{cond1}) and Eq.(\ref{cond2}) are compatible if 
(for $w<-1/3$)
\beq
10^{21}10^{16/(1+w)}
(t_d/10^{34}{\rm yr})^{2/3(1+w)}
 < 10^{49/(1-w)}(t_d/10^{34}{\rm yr})^{2/3(1-w)}.
\eeq
Taking the logarithm, the above condition is rewritten  as
\beq
{4w\over 3(1-w^2)}\log(t_d/10^{34}{\rm yr})+{33+65w\over 1-w^2}-21>0.
\label{ratio}
\eeq
It is easily found that the left-hand-side of Eq.(\ref{ratio}) is an 
increasing function of $w$ (for $-1<w<-1/3$) and negative as 
long as 
\beq
w<-1/3
\eeq
for $t_d>10^{33}{\rm yr}$\footnote{The condition would apply to the 
equation of state at the time of formation.  For dark energy which has the present 
equation of state $w<-1/3$ temporarily and then would increase toward $w>-1/3$ (for example, 
quintessence axion \cite{axion}), the formation of positronium would be possible.}
Hence positronium will not form in the universe dominated 
by dark energy with $w<-1/3$.\footnote{This does not mean that positronium will never 
form in the universe with $w<-1/3$. Although some small bit of activity always takes place, 
it will be exponentially suppressed.\cite{busha}}
We will restrict ourselves to the case with 
$w>-1/3$ hereafter. 

The formation of positronium through collisional recombination is studied in appendix. 
Here we only give the results: the typical time scale for the formation is 
$\sim 10^{83}{\rm yr}(t_d/10^{34}{\rm yr})$ and the size of positronium is 
$\sim 10^{45}{\rm Mpc}$ for $w=-0.1$ (see Fig.3 and Fig.4).

\section{Positronium Decay and Annihilation}

We have seen that positronium will be formed if $w>-1/3$. 
Positronium, once formed long after nucleon decay,  will decay toward 
ground state and then 
pair-annihilate. One can use the corresponding principle to calculate 
the time within which the positronium will spiral into the ground state by 
dipole radiation (for large $n$) \cite{ll,pm}
\beq
t_{spiral}\sim m^{-1}e^{-10}n^6.
\eeq
{}From Eq.(\ref{colleq}) in Appendix up to $n_{max}$, the fraction of free $\epm$'s 
evolves as 
\beq
y_e\propto a^{-3(2-\alpha)/5\alpha},
\eeq
where the expansion law is $a\propto t^{\alpha}$, $\alpha\equiv 2/3(1+w)$. 
Then a lifetime of decay to the ground state is
\beq
t_{spiral} \propto n^6_{max}\propto  y_e^{-1}a^3\propto 
a^{6(1+2\alpha)/5\alpha}.
\eeq
Expressing $y_e$ in terms of $t_{spiral}$ gives the logarithmic 
annihilation rate coefficient \cite{pm}
\beq
{\beta \over t}={2-\alpha\over 2+4\alpha}{1\over t}=
{2+3w\over 7+3w} {1\over t}.
\eeq
Thus, after the formation of positronium,  $\epm$'s density $\rho_e$ and 
radiation density $\rho_r$ evolve according to 
\beqa
\dot\rho_e&=&-3H\rho_e-{\beta t^{-1}}\rho_e \\
\dot\rho_r&=&-4H\rho_r+{\beta t^{-1}}\rho_e. \\
\eeqa
Assuming that $\rho_e=\rho_{ed}$ and $\rho_r=\rho_{rd}$ at $t=t_d$, 
the solutions are
\beqa
&&\rho_e=\rho_{ed}(t/t_d)^{-(3\alpha+\beta)}\\
&&\rho_r={\beta\rho_{ed}\over \alpha-\beta}
\left((t/t_d)^{-(3\alpha+\beta)}-(t/t_d)^{-4\alpha}\right)
+\rho_{rd}(t/t_d)^{-4\alpha}.
\eeqa
The asymptotic ratio of $\rho_e$ to $\rho_r$ 
is given by 
\beq
{\rho_e\over \rho_r}={\alpha-\beta\over \beta}={8-9w-9w^2\over 3(2+3w)(1+w)}.
\eeq
In Fig. 2, the ratio is shown as a function of $w$. 
The ratio is greater than $4/3$ for $w<0$. 
Hence, the radiation by $\epm$ pair annihilation cannot dominate over 
$\epm$. This is because in the universe with dark energy 
the expansion rate is higher and radiation suffers from much larger redshift. 

\begin{figure}
\includegraphics[width=\hsize]{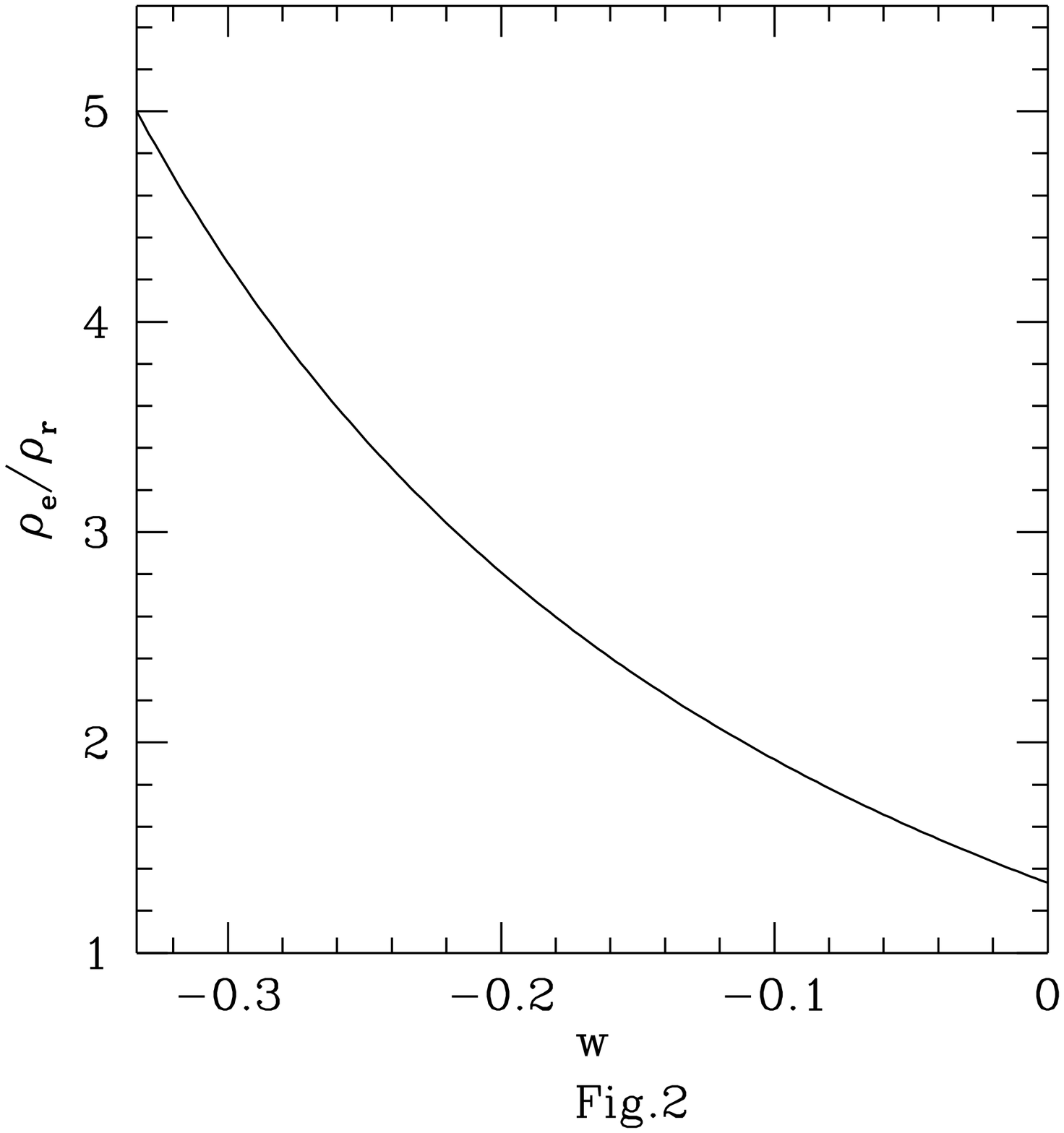}
\caption{
The ratio of $\epm$'s density to radiation density.}
\end{figure}

\begin{table}
  \begin{center}
  \setlength{\tabcolsep}{3pt}
  \begin{tabular}{c|ccccr} \hline
now & Dark Energy& Dark Matter& Baryon & Radiation& $e^\pm$\\
$10^{34}$yr & Dark Energy& Dark Matter& Radiation & Baryon& $e^\pm$ \\
$10^{38}$yr & Dark Energy& Dark Matter & $e^\pm$ & Radiation& \\
$10^{83}$yr & Dark Energy& Dark Matter & $e^\pm$/Ps & Radiation& \\
 \hline
  \end{tabular}
  \end{center}
\caption{
The energy content of the universe assuming $t_d=10^{34}$ years. PS denotes positronium.
}
\end{table}

\section{Summary}

We have investigated the matter content in the long-term future of 
the universe. After the proton decay, radiation becomes dominant over 
baryon and $\epm$ for a while and then is followed by them again but it will 
not dominate over dark matter.

We have also investigated the possibility of formation of positronium 
after proton decay. We have found that positronium will not be 
formed in the universe if the equation of 
state of dark energy $w$ is less than $-1/3$. 
Positronium will be formed after $10^{83}$ years 
if the equation of state of dark energy is $w>-1/3$, and radiation 
will be produced by positronium decay. However,  it will never 
dominate over matter due to the slow annihilation rate. The results are 
summerized in Table 1. 

\ack

This work supported in part by a Grant-in-Aid for Scientific 
Research (No.15740152,No.14340290) from the Japan Society for the Promotion of
Science.

\section*{Appendix: Collisional Recombination}

In this appendix, following Page and McKee \cite{pm}, we give the detailed analysis 
of positronium formation by collisional recombination.

As shown by Page and McKee \cite{pm}, 
radiative processes are unimportant for free $\epm$. Therefore  
positronium will form by collisional recombination,
\beq
 e^{-}+e^{+}+\epm\ra \ps_n+\epm,
\label{tran1}
\eeq
where $\ps_n$ denotes positronium with principal quantum number $n$. Once 
formed, positronium will tend to decay toward the ground state or at least 
an $S$ state and then annihilate. At first, the transitions will be induced 
mainly by collisions (for large $n$ states),
\beqa
&&\ps_n+\epm \ra \ps_{n'}+\epm\label{tran2}\\
&&\ps_n+\ps_{n''}\ra\ps_{n'}+\ps_{n'''}.
\eeqa

Let $N_e$ and $N_n$ be the number densities of free $\epm$'s and $\ps_n$'s.
We introduce the fractions of free $\epm$'s and $\ps_n$'s
\beq
y_e=N_e/N,~~~y_n=N_n/N,
\eeq
where $N$ is the total number density of all $\epm$'s.

The transitions Eq.(\ref{tran1}) and Eq.(\ref{tran2}) and their inverses 
occur at rates proportional to the densities of the participating particles, 
The rate is given by
\beq
{dy_n\over dt}={1\over 4}N^2y_e^3C_{i,n}-Ny_eC_{n,i}y_n+
Ny_e\sum_{n'}\left(\alpha(n',n)y_{n'}-\alpha(n,n')y_n\right),
\label{evolv}
\eeq
where $C_{i,n}$ is the collisional recombination rate coefficient to 
the $n$th level, $C_{n,i}$ is the collisional ionization rate coefficient 
from the $n$th level, and $\alpha(n,n')$ is the collisional transition 
rate coefficient. They are defined by
\beqa
&&C_{n,i}=\langle\sigma^{ion}(n)v \rangle \label{collision}\\
&&\alpha(n,n')=\langle\sigma(n,n')v\rangle,\label{bound-bound}
\eeqa
where $\sigma^{ion}(n)$ is the cross section for ionization from 
the $n$th level, $\sigma(n,n')$ is the cross section for transitions from
the $n$th to the $n'$th level, and the averages are taken over the Maxwellian 
distribution of the relative velocities. By detailed balance 
$C_{i,n}$ is given by 
\beq
C_{i,n}=8n^2(\pi/mT)^{3/2}\exp(me^4/4n^2T)C_{n,i},\label{detailed}
\eeq
where $T$ is the $\epm$ temperature, $T\sim \gamma^2m(a_d/a)^2$. Note that 
the binding energy of positronium is $me^4/4n^2T$. 

The ionization cross section of an atom is given in \cite{pr}
\beq
\sigma^{ion}={\pi e^4(5E+2I_n)(E-I_n)\over 3E^2I_n(E+3I_n)},
\eeq
where $E$ is the energy of the incident electron and $I_n$ is the 
ionization energy of the atom ($I_n=me^4/4n^2T$ for positronium). 
However, for our crude purposes, a rough expression at high and low 
energies is adequate:
\beq
\sigma^{ion}\sim {e^4(E-I_n)\over E^2I_n}.
\eeq
In terms of the ratio of ionization energy to temperature $x_n=I_n/T$, 
$C_{n,i}$  and $C_{i,n}$ are given by using Eq.(\ref{collision}) and 
Eq.(\ref{detailed})
\beqa
&&C_{n,i}\sim n^2m^{-3/2}T^{-1/2}(1+x_n)^{-1}\exp(-x_n),\\
&&C_{i,n}\sim n^4m^{-3}T^{-2}(1+x_n)^{-1}.
\eeqa

The collisional cross sections for transitions between highly excited 
levels of hydrogen have been considered in several papers \cite{pr,gee}
(see \cite{pm} for a list of references). An interpolation formula 
in \cite{gee}) gives the cross section which is accurate to 15\% for
$I_n\siml E\ll m, e^2n^{-1}\siml v \ll 1$. If we drop numerical factors 
and logarithms, then for $1\ll s\equiv n'-n\ll n$ the cross section 
become
\beqa
&&\sigma(n,n+s)\sim n^4s^{-3}m^{-1}E^{-1}, ~~~me^4n^{-1}s^{-1}<E<m\\
&&\sigma(n,n+s)\sim n^6s^{-1}e^{-8}m^{-3}E,~~~me^4n^{-2}<E<me^4n^{-1}s^{-1}.
\eeqa
{}From Eq.(\ref{bound-bound}), we obtain
\beqa
&&\alpha(n,n+s)\sim n^4s^{-3}m^{-3/2}T^{-1/2}, ~~~me^4n^{-1}s^{-1}<E<m  
\label{alpha1}\\
&&\alpha(n,n+s)\sim n^6s^{-1}e^{-8}m^{-7/2}T^{3/2},
~~~me^4n^{-2}<E<me^4n^{-1}s^{-1}.
\label{alpha2}
\eeqa
By detailed balance, the deexcitation rate coefficients are give by
\beq
\alpha(n+s,n)=(1+s/n)^{-2}\exp(x_n-x_{n+s})\alpha(n,n+s).
\label{detail2}
\eeq

It is apparent that from Eq.(\ref{alpha1}) and Eq.(\ref{alpha2}) that 
for $T>me^4n^{-2}$ the dominant bound-bound transitions are for relatively 
small changes $s$ in the level $n$. Hence if one views 
\beqa
&&\alpha_s=\alpha_s(n,t)\equiv \alpha(n,n+s)\\
&&y=y(n,t)\equiv y_n
\eeqa
as smooth functions of $n$ as well as of $t$, uses the detailed balanced 
relation Eq.(\ref{detail2}), and expands Eq.(\ref{evolv}) to second order in 
$s=n'-n$, one gets a diffusion equation
\beqa
{\partial y\over \partial t}&\simeq& {1\over 4}N^2y_e^3C_{i,n}-Ny_eC_{n,i}y
\nonumber\\
&&+{\partial\over \partial n}\left(Ny_e\sum_{s=1}^{n}\alpha_ss^2n^2\exp(x_n)
{\partial\over \partial n}\left(\exp(-x_n)yn^{-2}\right)\right).
\eeqa
Neglecting a $\ln n$ factor, the sum over $s$ gives
\beq
\sum_{s=1}^{n}\sim n^4m^{-3/2}T^{-1/2}, ~~~me^4n^{-2}<T<m.
\eeq
Now if we define the fraction of $\epm$'s in each individual positronium 
state as 
\beq
z\equiv n^{-2}y(n,t)/2,
\eeq
then for nonrelativistic temperatures well above the ionization temperature 
($T\ll m$ and $x_n\ll 1$), 
\beq
{\partial z\over \partial t}\sim {Ny_e\over m^{3/2}T^{1/2}}\left[
n^2\left({Ny_e^2\over m^{3/2}T^{3/2}}-z\right) +{1\over n^2}{
\partial\over \partial n}\left(n^6{\partial z\over \partial n}\right)\right].
\eeq

When $y_e\sim 1$, the population of positronium states with $x_n\ll 1$ 
increases toward the limiting value
\beqa
z_S(n,t)&\sim& Ny_e^2(mT)^{-3/2}\\
&\sim& N_dy_e^2(\gamma m)^{-3} \sim 10^{-48}10^{-48/(1+w)}y_e^2(t_d/10^{34}
{\rm yr})^{-2/(1+w)},
\eeqa
which is the Saha equilibrium value for $x_n\ll 1$. 
The population reaches this nearly stationary equilibrium only for levels 
in which the diffusion time is short compared with the expansion time $t$. 
Populations are only filled up to $z_S$ for
\beqa
n>n_{eq}(t)&\sim& y_e^{-1/2}N^{-1/2}(mT)^{3/4}t^{-1/2}\\
&\sim &10^{-10}10^{24/(1+w)}y_e^{-1/2}(t_d/10^{34}{\rm yr})^{(1-w)/(2+2w)}
(t/t_d)^{(1-3w)/(6+6w)}.
\eeqa

Since positronium must have a orbital size $r_n(=2m^{-1}e^{-2}n^2)$ less 
than the mean separation $(y_eN)^{-1/3}$ between free $\epm$'s,
\beqa
n<n_{max}(t)&\sim& m^{1/2}e(y_eN)^{-1/6}\\
&\sim& 10^{6}10^{8/(1+w)}y_e^{-1/6}(t_d/10^{34}{\rm yr})^{1/3(1+w)}
(t/t_d)^{1/3(1+w)}.
\eeqa
As the universe expands, $n_{max}(t)$ increases and more $\ps_n$ levels 
become available to be filled up to population $z_S$. 
The filling up of levels with $x_n<1$ decreases the fraction of free 
$\epm$'s by
\beq
dy_e\sim -z_Sn_{max}^2dn_{max}\sim -y_e^{3/2}a^{1/2}da/a_2^{3/2},
\eeq
where the time scale $a_2$ is
\beq
a_2/a_d\sim \gamma^2me^{-2}N_d^{-1/3}\sim 10^{21}10^{16/(1+w)}
(t_d/10^{34}{\rm yr})^{2/3(1+w)}.
\eeq
This is the same as $a_*$ given by Eq.(\ref{cond1}) since the binding 
energy, $I_n$,  is smaller than the average kinetic energy.

However, for level $x_n>1$ or
\beqa
n<n_{thermal}&=&(me^4/4T)^{1/2}\\
&\sim&e^2\gamma^{-1}(a/a_d)\sim 10^{-5}(t/t_d)^{2/3(1+w)},
\eeqa
the exponential increase in the Saha equilibrium value above $z_S$ 
(for small $x_n$) allows the value of $z$ to exceed $z_S$ if the 
collisional recombination rates are fast enough. Bound-bound transitions 
are less important in this case, and ionization becomes exponentially 
damped for $x_n>1$, so
\beq
{\partial z\over \partial t}\simeq {N^2y_e^3\over 8n^2}C_{i,n}\sim
{n^4N^2y_e^3\over m^4e^4T}\sim {n^4N_d^2y_e^3\over \gamma^2m^5e^4}(a/a_d)^{-4}.
\label{colleq}
\eeq
This implies that once the ionization becomes unimportant for levels $n$, 
that level will fill up to population
\beq
z\sim n^4N_d^2y_e^3\gamma^{-2}m^{-5}e^{-4}(a/a_d)^{-4}t.
\eeq
This in turn depletes the free $\epm$'s by
\beq
dy_e\sim -zn_{thermal}^2dn_{thermal}\sim y_e^3a^{2+3(1+w)/2}da/
a_1^{3+3(1+w)/2},
\eeq
with the time scale $a_1$, where
\beqa
a_1/a_d&\sim& [e^{-10}\gamma^9m^5N_d^{-2}H_0(a_0/a_d)^{3(1+w)/2}]^{2/(9+3w)}\\
&\sim& 10^{(296+104w)/(9+3w)/(1+w)}(t_d/10^{34}
{\rm yr})^{2(3-w)/(9+3w)/(1+w)}.
\eeqa
In Fig. 3, $t_1$ and $t_2$ are plotted as a function of $w$, 
where $t_1(t_2)$ is the time corresponding to $a_1(a_2)$. Typically
$t_1\sim 10^{83}{\rm yr}(t_d/10^{34}{\rm yr})$.

\begin{figure}
\includegraphics[width=\hsize]{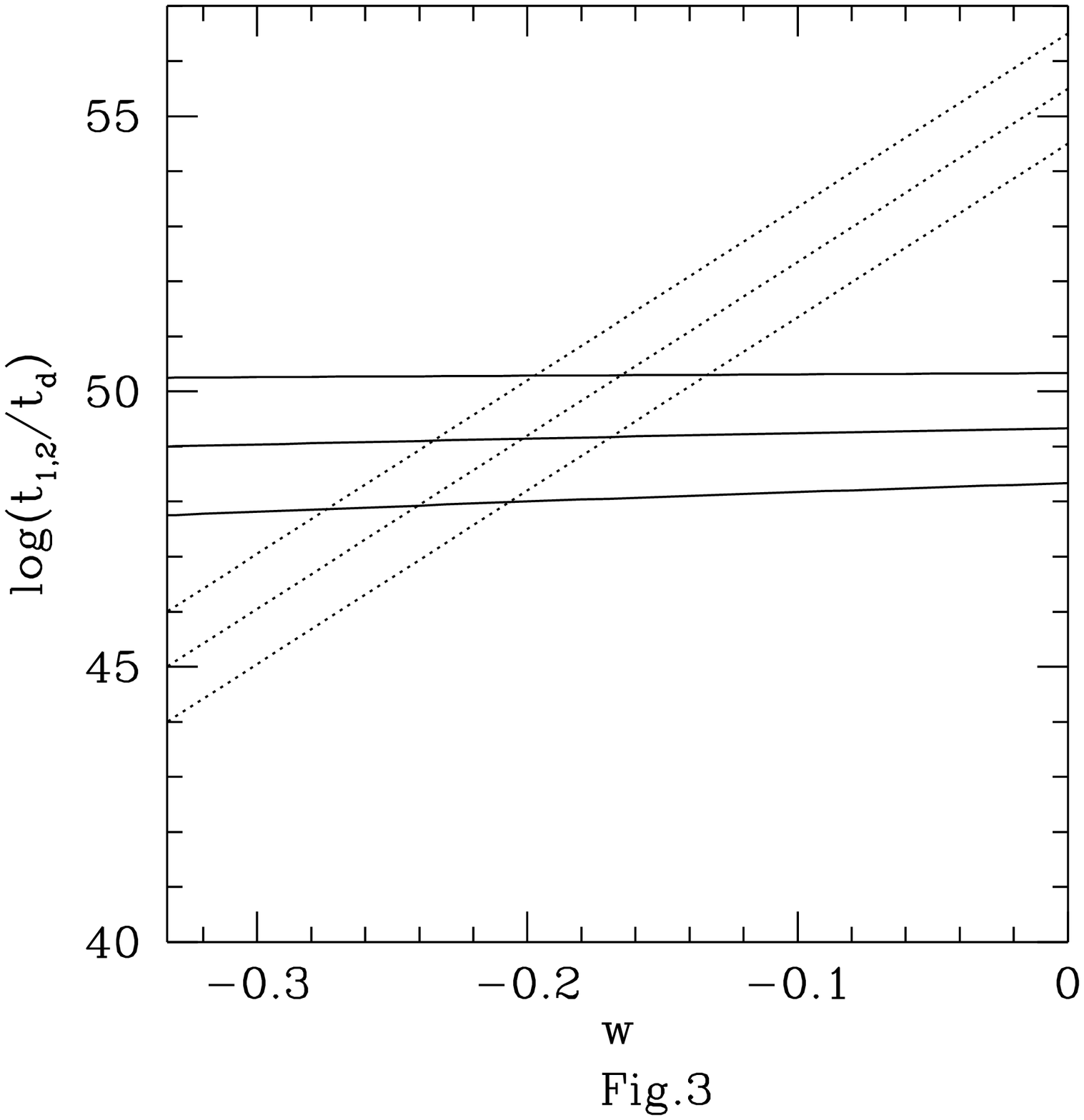}
\caption{
The formation times by collisional recombination
($t_1$: solid lines) and by bound-bound transitions ($t_2$: dashed lines)
as  functions of $w$. Three line are for $t_d=10^{33}{\rm yr},
10^{34}{\rm yr}, 10^{35}{\rm yr}$ from top to bottom.}
\end{figure}

For $-0.2\siml w< 0$, $t_1<t_2$,
Most $\epm$'s will bind 
around the time $t_1$, going into $\ps_n$ levels $n$ somewhat below
\beq
n_1=n_{thermal}(t_1)\sim 10^{-5}(a_1/a_d).
\eeq
These positronium states are very loosely bound, with radii
\beqa
r_n&\sim& n_1^2e^{-2}m^{-1}\\
 &\sim& 10^{-43}10^{(592+208w)/(9+3w)/(1+w)}(t_d/10^{34}
{\rm yr})^{4(3-w)/(9+3w)/(1+w)}{\rm Mpc}.
\eeqa
$r_n\sim 10^{30}$Mpc for $w=-0.1$. 

For $w\siml -0.2$, however, since the expansion rate of the universe 
becomes higher, collisional recombination is no longer effective, and 
hence $t_1>t_2$. The typical size of positronium is
\beqa
r&\sim& e^2\gamma^{-2}m^{-1}(a_2/a_d)^2\\
&\sim& 10^{-1}10^{32/(1+w)}(t_d/10^{34}{\rm yr})^{4/3(1+w)}{\rm Mpc}.
\eeqa
$r\sim 10^{45}$Mpc for $w=-0.3$. The results are shown in Fig.4.

\begin{figure}
\includegraphics[width=\hsize]{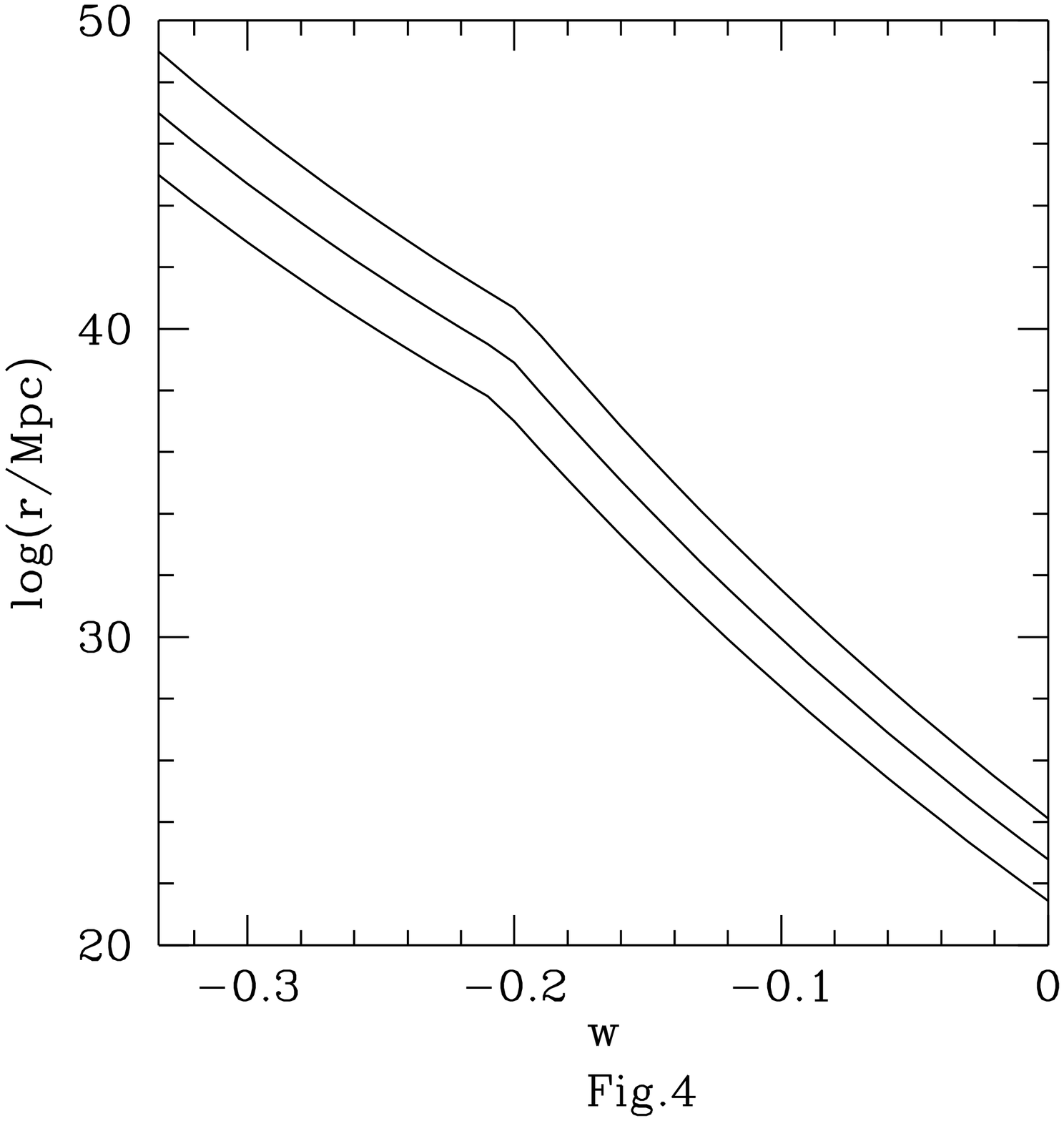}
\caption{
The size of positronium as a function of $w$. 
Three line are for $t_d=10^{33}{\rm yr},
10^{34}{\rm yr}, 10^{35}{\rm yr}$ from bottom to top.}
\end{figure}


\end{document}